\documentclass[english,twoside,a4paper,10pt]{article}

\usepackage[latin1]{inputenc}
\usepackage[T1]{fontenc}
\usepackage{dsfont}
\usepackage{amsmath}
\usepackage{amsfonts}
\usepackage{graphicx}
\usepackage{subfigure}
\usepackage{a4wide}
\usepackage{amssymb}
\usepackage{fancyhdr}
\usepackage{mathrsfs}
\usepackage[toc,page]{appendix}

\begin{document}

\title{\bf Viscous Fluids and Gauss-Bonnet  Modified  Gravity}
\author{Olesya Gorbunova\footnote{Also at Tomsk State Pedagogical University. E-mail address:
olesya@science.unitn.it, gorbunovaog@tspu.edu.ru
} and Lorenzo Sebastiani\footnote{E-mail address:
l.sebastiani@science.unitn.it
}\\
\\
\begin{small}
Dipartimento di Fisica, Universit\`a di Trento \end{small}\\
\begin{small}and Istituto Nazionale di Fisica Nucleare\end{small}\\ 
\begin{small}Gruppo Collegato di Trento, Italia
\end{small}\\
}
\date{}

\maketitle

\def\thesection{\Roman{section}}
\def\theequation{\Roman{section}.\arabic{equation}}


\begin{abstract}

We study effects of cosmic fluids on finite-time future singularities in modified $f(R,G)$-gravity,
where $R$ and $G$ are the Ricci scalar and the Gauss-Bonnet invariant,
respectively. We consider the fluid equation of state in the general form, $\omega=\omega(\rho)$, and we suppose the existence of a bulk viscosity. We investigate quintessence region ($\omega>-1$) and phantom region ($\omega<-1$) and the possibility to change or avoid the singularities in $f(R,G)$-gravity. Finally, we study the inclusion of quantum effects in large curvature regime.

\end{abstract}



\section{Introduction}

\paragraph*{} Recent observational data imply that the current expansion of the universe
is accelerating~\cite{WMAP, SN1}. This is the so called Dark Energy issue.
There exist several descriptions of  this fact. Among them, the simplest one is the introduction of small positive 
Cosmological Constant in the framework of General Relativity (GR), the so called $\Lambda$-CDM model. 
A generalization of this simple modification of GR consists in considering  modified gravitational theories, 
 in which the action is described by the Ricci scalar
$R$ plus an arbitrary function $f(R)$ of $R$ (for reviews, see~\cite{f(R)-gravity}). 

Alternatively, accelerating FRW universe may be  described by cosmological quintessence/ phantom scalar dark energy or
GR plus  fluid, satisfying suitable equation of state. It is also well-known that any of such dark energy-models
maybe represented as an  effective fluid with corresponding characteristics.

In this paper, we would like to investigate  modified $f(R,G)$ theories of gravity, where $G$ is the Gauss-Bonnet 
invariant. These models are generalization of $f(R)$ models, and they mainly are  inspired by (super)string theories ~\cite{F(G)-gravity, Monica}. 

It is also well-know that  many of such modified gravity models 
bring the future universe evolution to finite-time singularity \cite{Odintsov, miodue}.
The classification of the (four) finite-time future singularities has been made
in Ref.\cite{Nojiri:2005sx}. Some of these types future singularities are softer than other and not all physical quantities (scale factor, effective energy density and pressure) necessarly diverge at this finite future time.

The presence of finite-time future singularities may cause serious problems in the black holes or stellar astrophysics\cite{Maeda}. Thus, it is of some interest to understand if any natural scenario to cure such singularities exists. 

In the present paper, we generalize the results obtained in
Ref.~\cite{Odintsov} and \cite{miodue}, and
we explore the role of viscous fluids \cite{Alessia} within this class of modified $f(R,G)$ gravity models, investigating
 how the singularities may change or disappear, due to the contribution of these viscous fluids. 

The paper is organized as follows.
In Sec.~II,
we present the model of $f(R,G)$-gravity and the equation of state for generic viscous fluid. Hence, we
show the gravitational field equations and the energy conservation law of fluid.
In Sec.~III,
we explain the four types of the finite-time future
singularities.
Next, in Sec.~IV, we study the behaviour of fluids with constant thermodinamical parameter $\omega$
on finite-time future singularities.
Moreover, in Sec.~V, we explore the singularities in
the presence of a generic class of fluids with $\omega$ non constant.
In Sec.~V, we consider quantum effects related to singularities in the realistic Hu-Sawicki Model of $f(R)$-modified gravity. 
Finally, conclusions are given in Sec.\ VI, and in the Appendix we discuss the energy conditions for wich the singularities appear.

We use units of $k_\mathrm{B} = c = \hbar = 1$ and denote the gravitational constant $G_{N}$. 

\setcounter{equation}{0}

\section{Formalism}

\paragraph*{} We consider the simple action of $f(R,G)$-gravity:
\begin{equation}
S=\int d^{4}x\sqrt{-g}\left[\frac{R+f(R,G)}{16\pi G_{N}}+\mathcal{L}\right]\,,\label{azione}
\end{equation}
where $g$ is the determinant of the metric tensor $g_{\mu\nu}$, and $\mathcal{L}$ is the fluid Lagrangian, and $f(R,G)$ is a function of the Ricci scalar $R$ and the Gauss-Bonnet invariant $G=R^{2}-4R_{\mu\nu}R^{\mu\nu}+R_{\mu\nu\xi\sigma}R^{\mu\nu\xi\sigma}$ ($R_{\mu\nu}$ and $R_{\mu\nu\xi\sigma}$ are the Ricci tensor and the Riemann tensor, respectively). The general flat FRW space-time is described by the metric:
\begin{equation}
ds^{2}=-dt^{2}+a^{2}(t)d\mathbf{x}^{2}\,,
\end{equation}
where $a(t)$ is the scale factor of the universe. The Hubble parameter is $H=\dot{a}/a$ (the point denotes the derivative with respect to the cosmic time $t$) and for the above metric, the scalar curvature and the Gauss-Bonnet read:
\begin{equation}
 R= 6\left(2H^{2}+\dot{H}\right)\label{R}\,,
\end{equation}
\begin{equation}
 G=24 H^{2}\left(H^{2}+\dot{H}\right)\label{G}\,.
\end{equation}

The scalar expansion is $\nabla_{\mu}u^{\mu}=3 H$, $u^{\mu}$ being the four-velocity of the cosmic fluid introduced in Eq.(\ref{azione}). We assume the fluid equation of state in the form:
\begin{equation}
p=\omega(\rho)\rho-3 H\zeta(\rho)\,,\label{eq.state}
\end{equation}
where $p$ and $\rho$ are the pressure and energy density of fluid, respectively. The thermodynamical variable $\omega(\rho)$ is an arbitrary function of the density $\rho$. $\zeta(\rho)$ is the bulk viscosity and in general it depends on $\rho$. On thermodynamical grounds, in order to have the positive sign of the entropy change in an irreversible process, $\zeta(\rho)$ has to be a positive quantity, so we assume $\zeta(\rho)>0$. For the stress-energy tensor $T_{\mu\nu}$, one has :
\begin{equation}
T_{\mu\nu}=\rho u_{\mu}u_{\nu}+\left(\omega(\rho)\rho-3H\zeta(\rho)\right)(g_{\mu\nu}+u_{\mu}u_{\nu})\,. 
\end{equation}

The equations of motion (EOM) are derived from Eq.(\ref{azione}) by using the variational principle\cite{Monica}: 
\begin{equation}
\rho_{G}+\rho=\frac{3}{8\pi G_{N}}H^{2}\label{EOM1}\,,
\end{equation}
\begin{equation}
p_{G}+p=-\frac{1}{8 \pi G_{N}}\left(2\dot{H}+3H^2\right)\label{EOM2}\,.
\end{equation}
The part of modified gravity is formally included into the modified energy density $\rho_{G}$ and the modified pressure $p_{G}$ as follows:
\begin{equation}
\rho_{G}=-\frac{1}{16\pi G_{N}}\Bigl\{24H^{3}\dot{f}'_{G}+6H^{2}f'_{R}+6H\dot{f}'_{R}+(f-Rf'_{R}-Gf'_{G})\Bigr\}\label{rhoG}\,,
\end{equation}
\begin{eqnarray}
p_{G}&=&\frac{1}{16\pi G_{N}}\Bigl\{8H^{2}\ddot{f}'_{G}+2\ddot{f}'_{R}+4H\dot{f}'_{R}+16H\dot{f}'_{G}(\dot{H}+H^{2})+f'_{R}(4\dot{H}+6H^{2})+\nonumber \\ \nonumber \\ & & \phantom{spacespa}(f-R\dot{f}'_{R}-Gf'_{G})\Bigr\}\label{pG}\,.  
\end{eqnarray}
Here, we have used the following expressions:
\begin{equation}
f'_{R}=\frac{\partial f(R,G)}{\partial R}\,,\phantom{spacespacespace}f'_{G}=\frac{\partial f(R,G)}{\partial G}\,. 
\end{equation}
Note that the effective energy density of the universe is $\rho_{eff}=\rho_{G}+\rho$ and the effective pressure is $p_{eff}=p_{G}+p$, whereas for the effective equation of state, we obtain:
\begin{equation}
\omega_{eff}\equiv \frac{p_{eff}}{\rho_{eff}}= -1-\frac{2\dot{H}}{3H^{2}}\,. \label{Lorenzobis}
\end{equation}

The fluid energy conservation law is a consequence of the EOM (\ref{EOM1})-(\ref{EOM2}):
\begin{equation}
\dot{\rho}+3H\rho(1+\omega(\rho))=9H^{2}\zeta(\rho) \label{conservationlaw}\,,
\end{equation}
and, in general, we write $\zeta(\rho)$ like a function of $H$, $\zeta(H)$.

\setcounter{equation}{0}

\section{Future-time singularities}

\paragraph*{} We are interested in the modified $f(R,G)$-models that produce some type of finite-time future singularities (see, for instance, \cite{Odintsov},\cite{miodue}), giving by the Hubble parameter:
\begin{equation}
H=\frac{h}{(t_{0}-t)^{\beta}}+H_{0}\label{Hsingular}\,,
\end{equation}
where $h$, $H_{0}$ and $t_{0}$ are positive constants and $t<t_{0}$ because it should be for expanding universe. $\beta$ is a positive constant or a negative non-integer number, so that, when $t$ is close to $t_{0}$, $H$ or some derivative of $H$ and therefore the curvature become singular. $H_{0}$ could be relevant in the limit $t\rightarrow t_{0}$ only when $\beta<0$, and we can assume $H_{0}=0$ when $\beta>0$.

Note that such choice of Hubble parameter corresponds to accelerated universe, because on the singular solution of Eq.(\ref{Hsingular}) the strong energy condition is violated (see the Appendix).

If $\beta=1$, Eq.(\ref{Hsingular}) with $H_{0}=0$ implies that the scale factor $a(t)$ behaves as:
\begin{equation}
a(t)=\frac{a_{0}}{(t_{0}-t)^{h}}\label{a}\,.
\end{equation}

If $\beta>0$ and $\beta\neq 1$, Eq.(\ref{Hsingular}) with $H_{0}=0$ yields:
\begin{equation}
a(t)=a_{0}e^{\frac{h(t_{0}-t)^{1-\beta}}{\beta-1}}\label{b}\,.
\end{equation}

In the end, if $\beta<0$, we have to take into account the constant $H_{0}\geqslant 0$ in Eq.(\ref{Hsingular}), and the general form of $a(t)$ is given by:
\begin{equation}
a(t)=a_{0} e^{-(t_{0}-t)\left(H_{0}-\frac{h(t_{0}-t)^{-\beta}}{\beta-1}\right)}\label{c}\,.
\end{equation}
Here, $a_{0}$($\neq 0$) and in what follows $\rho_{0}$ are positive constants.\\
\\
The finite-time future singularities can be classified in the following way\cite{Nojiri:2005sx}:
\begin{itemize}
\item Type I (Big Rip): for $t\rightarrow t_{0}$, $a(t)\rightarrow\infty$,
$\rho_\mathrm{{eff}}\rightarrow\infty$ and
$|p_\mathrm{{eff}}|\rightarrow\infty$. The case in which
$\rho_\mathrm{{eff}}$ and $p_\mathrm{{eff}}$ are finite at $t_{0}$ is also
included.
It corresponds to $\beta=1$ and $\beta>1$.
\item Type II (sudden):
for $t\rightarrow t_{0}$, $a(t)\rightarrow a_{0}$,
$\rho_\mathrm{{eff}}\rightarrow\rho_{0}$ and $|p_\mathrm{{eff}}|
\rightarrow\infty$.
It corresponds to $-1<\beta<0$.
\item Type III: for $t\rightarrow t_{0}$, $a(t)\rightarrow a_{0}$,
$\rho_\mathrm{{eff}}\rightarrow\infty$ and
$|p_\mathrm{{eff}}|\rightarrow\infty$.
It corresponds to $0<\beta<1$.
\item Type IV: for $t\rightarrow t_{0}$, $a(t)\rightarrow a_{0}$,
$\rho_\mathrm{{eff}}\rightarrow 0$, $|p_\mathrm{{eff}}|
\rightarrow 0$
and higher derivatives of $H$ diverge.
The case in which $\rho$ and/or $p$ tend to finite values is
also included. It corresponds to
$\beta<-1$ but $\beta$ is not any integer number.
\end{itemize}

We note that in the present paper, we call singularities for $\beta=1$ and those for $\beta>1$ as the "Big Rip" singularities and the "Type I" singularities, respectively.\\

In the absence of fluids, the effective density and pressure of the universe are given by modified gravity, so that $\rho_{eff}=\rho_{G}$ and $p_{eff}=p_{G}$.
A $f(R,G)$-model shows some types of singularity, if Eq.(\ref{Hsingular}) is a solution of the EOM, namely:  
\begin{equation}
\rho_{G}= \frac{3}{8\pi G_{N}}\Bigl\{\frac{h^{2}}{(t_{0}-t)^{2\beta}}+\frac{2hH_{0}}{(t_{0}-t)^{\beta}}+H_{0}^2\Bigr\}\label{EOM1bis}\,,
\end{equation}
\begin{equation}
p_{G}= -\frac{1}{8 \pi G_{N}}\Bigl\{\frac{2h\beta}{(t_{0}-t)^{\beta+1}}+\frac{3h^{2}}{(t_{0}-t)^{2\beta}}+\frac{6hH_{0}}{(t_{0}-t)^{\beta}}+3H_{0}^{2}\Bigr\}\label{EOM2bis}\,.
\end{equation}
 
We want to see how the presence of viscous fluids influences the behaviour of singular $f(R,G)$-models (i.e. models that in absence of fluids produce some singularities and satisfy Eq.(\ref{EOM1bis})-(\ref{EOM2bis})). We will check the solutions of the fluid energy density when $H$ is singular by using the energy conservation law of Eq.(\ref{conservationlaw}), and we will see how changes the effective density (and, as a consequence, the effective pressure) of the universe, and if the singularities are still realized. In particular, we are interested in the quintessence ($-1<\omega<0$) and phantom ($\omega<-1$) region (referring to an ideal fluid), and we assume that the contribute of ordinary matter and radiation in expanding universe is too small with respect to the modified gravity.

We investigate the cases of $\omega$ constant and $\omega$ dependent on energy density $\rho$.

\setcounter{equation}{0}

\section{$\omega$ constant}

\paragraph*{} Let us start considering the simple case when $\omega$ is a constant. We take different choices of bulk viscosity $\zeta$.

\subsection*{A. Non-viscous case}

\paragraph*{} In the non-viscous case $\zeta=0$ (perfect fluid), the solution of Eq.(\ref{conservationlaw}) assumes the classical form:
\begin{equation}
\rho=\rho_{0}a^{-3(1+\omega)}\,,
\end{equation}
where $\rho_{0}$ is a positive constant and $a$ is the scale factor. As a consequence, on the sigular form of $H$ of Eq.(\ref{Hsingular}), $\rho$ behaves as (see Eq.(\ref{a})-(\ref{b})-(\ref{c})):
\begin{equation}
\rho= \rho_{0}(t_{0}-t)^{3h(1+\omega)}\,,\phantom{spacespacespace}\text{when $\beta=1$}\,,\label{tric}
\end{equation}
\begin{equation}
\rho=\rho_{0}e^{\frac{3h(1+\omega)(t_{0}-t)^{1-\beta}}{1-\beta}}\,,\phantom{spacespa}\text{when $\beta>0$}\,,\beta\neq 1\,,\label{trac}
\end{equation}
\begin{equation}
\rho\simeq\rho_{0}e^{3(1+\omega)(t_{0}-t)\left(H_{0}-\frac{h(t_{0}-t)^{-\beta}}{\beta-1}\right)}\,,\phantom{spac}\text{when }\beta<0\,.\label{truc}
\end{equation}

For $\beta=1$ (Big Rip) and $\beta>1$ (Type I singularity), $\rho$ grows up and becomes relevant when $t$ is close to $t_{0}$ only if $\omega<-1$. It means that phantom fluids increase the effective density and pressure of the universe in the case of Big Rip and Type I singularities, whereas quintessence fluids ($\omega>-1$) become negligible and do not influence the asymptotic behaviour of $f(R,G)$ models that realize this kind of singularities.

In Einstein's gravity ($f(R,G)=0$), Eq.(\ref{tric}) and Eq.(\ref{EOM1}) admit the solution:
\begin{equation}
H=-\frac{2}{3(1+\omega)}\frac{1}{(t_{0}-t)}\,,\label{oraetlabora}
\end{equation}
and we can see that the phantom fluid produces the Big Rip for $H=h'/(t_{0}-t)$, where $h'=-2/3(1+\omega)$.  

In general, in $f(R,G)$-modified gravity, in the presence of phantom fluid, the asymptotically Big Rip singularity could appear if $\rho_{G}$ in Eq.(\ref{rhoG}) diverges less than $H^{2}$ ($\sim (t_{0}-t)^{-2}$) on the singular solution of Eq.(\ref{oraetlabora}), namely the modified gravity becomes negligible with respect to the fluid energy density in Eq.(\ref{EOM1}). On the other hand, if a $f(R,G)$ model realizes the Big Rip for a certain value of $h$, the fluid energy density $\rho$ of Eq.(\ref{tric}) becomes negligible on this singular solution if $\omega>-(1+2/3h)$, because it diverges less than $H^{2}$ in Eq.(\ref{EOM1}).

When $\beta>1$, the energy density $\rho$ of phantom fluid diverges exponentially in Eq.(\ref{trac}), so that the EOM (\ref{EOM1})-(\ref{EOM2}) become inconsistent and the Type I singularity is not realized in $f(R,G)$-gravity.
 
When $0<\beta<1$, $\rho$ tends to $\rho_{0}$ with time in Eq.(\ref{trac}), and it is asymptotically negligible with respect to $H^{2}$ ($\sim (t_{0}-t)^{-2\beta}$). In this case, a $f(R,G)$-model realizing Type III singularity, is not influenced by perfect fluids on this kind of singularity.

For Type II and IV singular models ($\beta<0$), the presence of quintessence or phantom fluids can make the singularities worse. Note that $H^{2}$ of Type II and IV singularities tends to the constant $3H_{0}^{2}/8\pi G_{N}$ like $\sim(t_{0}-t)^{-\beta}$, while $\rho$ in Eq.(\ref{truc}), after the developing of the exponential function in power series, tends to $\rho_{0}$ like $\sim(1+\omega)(t_{0}-t)$. 

In the case of $-1<\beta<0$, a large value of energy density $\rho_{0}$ becomes relevant in Eq.(\ref{EOM1}) and could change the numerical value of $H_{0}$ for which the singularity appears in $f(R,G)$-gravity, but does not necessarly avoid  the singularity. 

In the case of $\beta<-1$, the dynamical behaviour of Eq.(\ref{EOM1}) could become inconsistent, because $\rho$ behaves as $(t_{0}-t)$ and it is larger than the time-dependent part of $H^{2}$ ($\sim(t_{0}-t)^{-\beta}$). In particular, the Type IV singularities with $|\beta|>>1$ are very difficult to realize in the presence of phantom or quintessence fluids.\\
\\
Examples:
\begin{itemize}
\item In the model $f(G)=-\alpha\sqrt{G}$, where $\alpha$ is a positive constant, the Type I singularity or the Big Rip for some values of $h>1$ could occur\cite{miodue}. If we add a phantom fluid ($\omega<-1$), the Type I singularity is avoided, while the Big Rip could still appear.\\ 
If $\omega<-5/3$ (namely, $\omega<-(1+2/3h)$ for any value of $h>1$), the fluid energy density of Eq.(\ref{tric}) grows up faster than $H^{2}$ when $h>1$, and this kind of Big Rip is not realized. On the other hand, the phantom fluid could produce the Big Rip for some value of $0<h'<1$, when $h'=-2/3(1+\omega)$ like in Eq.(\ref{oraetlabora}). However, it is possible to verify, by using Eq.(\ref{rhoG}), that $\rho_{G}$ of this model, when $0<h'<1$, diverges still like $H^2$, but is negative, so that, if its modulus is larger than the fluid density, the effective energy density of the universe becomes negative and the Big Rip is avoided, because Eq.(\ref{EOM1}) results inconsistent. It depends on the value of $\alpha$ parameter with respect to the fluid density $\rho_{0}$.\\

\item The model $f(R)=\alpha R^{\gamma}$, where $\alpha$ is a constant, could realize a Type II singularity when $\gamma<0$ or a Type IV singularity when $\gamma>2$ \cite{Odintsov}. In both cases we assume $H_{0}$ negligible in Eq.(\ref{Hsingular}).\\ 
The presence of quintessence or phantom fluids does not avoid the Type II singularity, because the numerical value of $H_{0}$ changes on the singular solution ($H_{0}=\sqrt{8\pi G_{N}\rho_{0}/3}$), but the dynamical behaviour of the modified function $f(R)$ keeps the same, due to the fact that $R$ tends to infinity in Eq.(\ref{R}), and is not influenced by the costant $H_{0}$. Moreover, if we use a phantom fluid, there is the possibility that the Type II singularity is changed into the Big Rip in the form of Eq.(\ref{oraetlabora}), because, when $H\sim (t_{0}-t)^{-1}$, it is easy to verify that $\rho_{G}$ of the model tends to zero, so that the fluid is dominant and makes the future singularity stronger.\\
When $\gamma>2$, the Type IV singularity could be avoided by phantom or quintessence fluids, especially if $\gamma$ is very close to two. This is because the model is singular for $\beta=\gamma/(2-\gamma)$, so that $|\beta|>>1$ if $\gamma$ is close to two. As a consequence, other future scenarios for the universe are possible. For example, if $\gamma=3$, the model admits an instable de Sitter solution with $R_{dS}=\sqrt{1/\alpha}$ \cite{mio}, or the phantom fluid may produce an accelerating phase. 

\item The model $f(G)=-\alpha G^{\gamma}$, where $\alpha>0$ and $\gamma>1$, shows the Type II singularity with $H_{0}$ negligible and $-1<\beta<-1/3$ in Eq.(\ref{Hsingular}). Now, the presence of phantom or quintessence fluids with large density $\rho_{0}$, avoids the Type II singularity. The value of $H_{0}$ and the dynamical behaviour of $f(G)$ change together, because in the case of $H_{0}=0$, when $-1<\beta<-1/3$, G tends to zero in Eq.(\ref{G}), but if $H_{0}$ is substantially different to zero, $G$ diverges to infinitive and Eq.(\ref{EOM1}) for this kind of model becomes inconsistent on the Type II singularity.

\end{itemize}

\subsection*{B. Constant viscosity}

\paragraph*{} Now, we introduce bulk viscosity in cosmic fluid. Note that viscous fluids belong to more general inhomogeneous EoS fluids introduced in Ref.\cite{Capozielloetal}.\\

Suppose to have the bulk viscosity equal to a constant, $\zeta=\zeta_{0}$. Eq.(\ref{conservationlaw}) yields:
\begin{equation}
\rho=\rho_{0}a^{-3(1+\omega)}+9\zeta_{0} a^{-3(1+\omega)}\int^{t} dt'a(t')^{1+3\omega}\dot{a}(t')^{2}\,.
\end{equation}

For the Big Rip ($\beta=1$), $\rho$ behaves as (see Eq.(\ref{a})):  
\begin{equation}
\rho=\rho_{0}(t_{0}-t)^{3h(1+\omega)}+\frac{9h^{2}\zeta_{0}}{(t_{0}-t)(1+3h+3h\omega)}\,.\label{zig}
\end{equation}

The bulk viscosity is not asymptotically relevant on the Big Rip solution, because the Hubble parameter $H^2$ diverges like $1/(t_{0}-t)^2$, while the viscosity part of $\rho$ diverges more slowly, like $1/(t_{0}-t)$, and in the EOM (\ref{EOM1})-(\ref{EOM2}) we can neglect it.

In the paper \cite{Alessia} is written the general exactly form of $H(t)$ for fluids with constant viscosity in absence of modified gravity ($f(R,G)=0$):
\begin{equation}
H(t)=\frac{\sqrt{24\pi G_{N}\rho_{0}} e^{(12\pi G_{N}\zeta_{0})t}}{3+\frac{3}{2}(1+\omega)\sqrt{24\pi G_{N}\rho_{0}}\frac{(e^{(12\pi G_{N}\zeta_{0})t}-1)}{12\pi G_{N} \zeta_{0}}}\,.\label{exp}
\end{equation}

$H(t)$ shows a finite-time future singularity when $t$ tends to $t_{0}$, where $t_{0}=(12\pi G_{N}\zeta)^{-1} \text{ln}(1-24\pi G_{N}\zeta_{0}/(1+\omega)\sqrt{24\pi G_{N}\rho_{0}})$. If we expand the exponential functions around $t_{0}$, we obtain:
\begin{equation}
H(t)\simeq-\frac{2}{3(1+\omega)}\frac{1}{(t_{0}-t)}+\frac{8\pi G_{N}\zeta_{0}}{1+\omega}+\mathcal{O}(t_{0}-t)\,,
\end{equation}
that corresponds to Eq.(\ref{Hsingular}) with $\beta=1$ (Big Rip), $h=-2/(3+3\omega)$, where $\omega<-1$, and $H_{0}=8\pi G_{N}\zeta_{0}/(1+\omega)$. The viscosity $\zeta_{0}$ is not relevant in the asymptotic limit of $H$ (here, $H_{0}$ is negative, but the first positive term of $H$ is much larger), and we recover Eq.(\ref{oraetlabora}), that is valid for phantom perfect fluids.\\

In order to study the effects of the viscosity on Type I, II, III and IV singular models, it is worth considering the asymptotic behaviour of the conservation law in Eq.(\ref{conservationlaw}). We require that the left part diverges like the right part on the singular solutions:
\begin{equation}
\dot{\rho}+3\rho(1+\omega)\left(\frac{h}{(t_{0}-t)^{\beta}}+H_{0}\right)\simeq \frac{9 h^{2}\zeta_{0}}{(t_{0}-t)^{2\beta}}+\frac{18 h H_{0}\zeta_{0}}{(t_{0}-t)^{\beta}}+9H_{0}^{2}\zeta_{0}\,,\label{claw}
\end{equation}
where we take $H_{0}=0$ if $\beta>0$. In what follows, we neglect the homogeneous solutions, already discussed in the previous chapter. 

The asymptotic solutions of Eq.(\ref{claw}) are:
\begin{equation}
\rho\simeq \frac{3 h \zeta_{0}}{(1+\omega)(t_{0}-t)^{\beta}}\,,\phantom{spac}\text{when }\beta>1\,,\label{zagzag}
\end{equation} 
\begin{equation}
\rho\simeq \frac{9\zeta_{0}h^{2}}{(2\beta-1)(t_{0}-t)^{2\beta-1}}\,,\phantom{space}\text{when }1>\beta>0\,,\label{otto}
\end{equation}
\begin{equation}
\rho\simeq\frac{9h H_{0}\zeta_{0}}{(\beta-1)(t_{0}-t)^{\beta-1}}+\frac{3H_{0}\zeta_{0}}{1+\omega}\,,\phantom{sp}\text{when }\beta<0\,,H_{0}\neq 0\,.\label{zag}
\end{equation}

In the first case ($\beta>1$), it is possible to see that $\rho$ diverges more slowly than $H^{2}$, so that viscous fluids do not influence the asymptotically behaviour of Type I singular models in Eq.(\ref{EOM1}), due to the constant viscosity.

Also in the second case ($0<\beta<1$), viscous fluids are asymptotically avoidable in the case of Type III singular models, because Eq.(\ref{otto}) diverges less than $H^{2}$. 
  
In the end, we consider fluids that tend to a non-negligible energy density when $\beta<0$. It automatically leads to $H_{0}\neq 0$ in Eq.(\ref{EOM1}) and $\rho$ behaves as in Eq.(\ref{zag}). Large bulk viscosity $\zeta_{0}$ becomes relevant in the EOM, determining the value of $H_{0}$ for which the singularity occurs in some model of $f(R,G)$-gravity. Moreover, if $\omega<-1$, the effective energy density (namely, $\rho_{G}+\rho$) could be negative and avoid the Type II and IV singularities for expanding universe ($H_{0}>0$).\\
\\
Example:
\begin{itemize}
\item The model $f(R)= \alpha(e^{-\lambda R^{2}}-1)$, where $\alpha$ and $\lambda$ are positive constants, shows the Type II singularity in the form of Eq.(\ref{Hsingular}) with $H_{0}=\sqrt{\alpha/6}$ and $\beta$ very close to $-1^{+}$ \cite{miodue}.\\
A fluid with $\omega>-1$ and constant viscosity $\zeta_{0}$ large with respect to $\alpha$, changes the value of $H_{0}$ ($H_{0}\simeq \sqrt{8\pi G_{N}\zeta_{0}/(1+\omega)}$), but the Type II singularity can still occur.\\
On the other hand, a fluid with $\omega<-1$ and $\zeta_{0}>>\alpha$, makes the singularity unphisical (the solution of Eq.(\ref{EOM1})-(\ref{zag}) leads to $H_{0}<0$ and the singularity is for contracting universe).

\end{itemize}

\subsection*{C. Viscosity proportional to H}

\paragraph*{} This is the case $\zeta=3H\tau$. As $\zeta$ is assumed to be positive, the constant $\tau$ has to be positive. Eq.(\ref{conservationlaw}) yields:
\begin{equation}
\rho=\rho_{0}a^{-3(1+\omega)}+27\tau a^{-3(1+\omega)}\int^{t} dt'a(t')^{3\omega}\dot{a}(t')^{3}\,.
\end{equation}

For the Big Rip ($\beta=0$), $\rho$ behaves as:  
\begin{equation}
\rho=\frac{27 h^{3}\tau}{(t_{0}-t)^{2}(2+3h+3h\omega)}\,,\label{zurp}
\end{equation}
and diverges like $H^{2}$ in Eq.(\ref{EOM1}), so that, in general, the Big Rip could still appear in the $f(R,G)$-models that realize this kind of singularity.

In absence of modified gravity ($f(R,G)=0$), Eq.(\ref{zurp}) and Eq.(\ref{EOM1}) admit the solution:
\begin{equation}
H=\frac{2}{(72\pi G_{N}\tau-3(1+\omega))}\frac{1}{(t_{0}-t)}\,,  \label{zip}
\end{equation}
and realize the Big Rip for $H=h'/(t_{0}-t)$, where $h'=2/(72\pi G_{N}\tau-3(1+\omega))$. $h'$ is positive if\cite{Alessia}:
\begin{equation}
(1+\omega)-24\pi G_{N}\tau<0\,.\label{eccoci}
\end{equation}

It means that, in $f(R,G)$-gravity, it could appear the Big Rip driven by phantom fluid or fluid in the quintessence region with sufficiently large bulk viscosity, if the modified gravity on the Big Rip solution is small in Eq.(\ref{EOM1}), namely $\rho_{G}$ diverges less than $H^{2}$ given by Eq.(\ref{zip}). On the other hand, if $(1+\omega)-24\pi G_{N}\tau>0$, the fluid does not realize the Big Rip for expanding universe.   

The other asymptotic solutions of Eq.(\ref{conservationlaw}) are:
\begin{equation}
\rho\simeq \frac{9 h^{2} \tau}{(1+\omega)(t_{0}-t)^{2\beta}}\,,\phantom{spa}\text{when }\beta>1\,,\label{uno}
\end{equation}
\begin{equation}
\rho\simeq \frac{27\tau h^{3}}{(3\beta-1)(t_{0}-t)^{3\beta-1}}\,,\phantom{spac}\text{when }0<\beta<1\,,\label{due}
\end{equation}
\begin{equation}
\rho\simeq\frac{27h H_{0}^{2}\tau}{(\beta-1)(t_{0}-t)^{\beta-1}}+\frac{9H_{0}^{2}\tau}{1+\omega}\,,\phantom{sp}\text{when }\beta<0\,,H_{0}\neq 0\,.\label{tre}
\end{equation}

For $\beta>1$, $\rho$ diverges like $H^2$ if $\omega>-1$. Thus the fluid could asymptotically produce the Type I singularity when the modified gravity is negligible or behaves as $\rho$ on this kind of solution. On the other hand, if $\omega<-1$, for large values of viscosity $\tau$, the Type I singularity could be avoided in $f(R,G)$-gravity, because the effective energy density of the universe becomes negative, making inconsistent Eq.(\ref{EOM1}).
 
When $0<\beta<1$, the fluid does not influence the Type III singular models and can be neglected on the singularity, because Eq.(\ref{due}) diverges less than $H^{2}$.
  
When $\beta<0$ and $H_{0}\neq 0$, the bulk viscosity can influence the singular solutions for large values of $\tau/(1+\omega)$ in Eq.(\ref{tre}). In particular, if $\omega<-1$, the effective energy density of the universe could become negative, avoiding Type II and IV singularities in $f(R,G)$-gravity.\\ 
\\
Examples:
\begin{itemize}
\item The model $f(R,G)=-\alpha(G/R)$, where $\alpha$ is a positive constant, shows the Type I singularity\cite{miodue}.\\
 A fluid with $\omega>-1$ and energy density in the form of Eq.(\ref{uno}), influences the feature of the singularity by changing some numerical value, but the Type I singularity is still realized. In addition, if $\tau$ is sufficiently large, Eq.(\ref{eccoci}) is satisfied and an other possible scenario is the Big Rip.\\
 If $\omega<-1$, large values of $\tau$ make negative the effective energy density of the universe on the Type I singularity, that could be changed into the Big Rip.

\item In the model $f(R,G)=\alpha(G/R)$, where $\alpha$ is a positive constant, the Type III, II and IV singularities could appear\cite{miodue}. The presence of fluids with $\omega<-1$ and large viscosity proportional to $H$, does not influence the Type III singularity, but could change the Types II and IV into the Big Rip, like in the previous example.    
\end{itemize}

\setcounter{equation}{0}

\section{$\omega$ not a constant}

\paragraph*{} In this general case, $\omega=\omega(\rho)$, we are interested in some simple solution of Eq.(\ref{conservationlaw}), when $H$ is singular and behaves as in Eq.(\ref{Hsingular}). We consider viscous fluid, whose thermodinamical parameter $\omega$ is given by:
\begin{equation}
\omega=A_{0}\rho^{\alpha-1}-1\label{omega}\,,
\end{equation}
where $A_{0}$($\neq 0$) and $\alpha$ are constants. When $\alpha=1$, we find the case when $\omega$ is a constant. Let us suppose the following form of bulk viscosity $\zeta$:
\begin{equation}
\zeta=(3 H)^{n}\tau\,.\label{viscosityinhomogeneus}
\end{equation}
Here, $n$ is a natural number and $\tau$ is a positive constant different to zero.

The energy conservation law leads:
\begin{equation}
\dot{\rho}+3 H A_{0}\rho^{\alpha}=9H^{2}(3H)^{n}\tau\label{super}\,,
\end{equation}
from which we may get the (asymptotic) solutions of the fluid energy density when $H$ is singular. 

In what follows, we consider several examples.\\

For the Big Rip singularity ($\beta=1$), some simple (asymptotic) solutions of Eq.(\ref{super}) are given by: 
\begin{equation}
\rho=\frac{3^{n+2}h^{n+2}\tau}{(n+1+3 h A_{0})(t_{0}-t)^{n+1}}\phantom{space}\text{for }\alpha=1\label{rhoin}\,,
\end{equation}
\begin{equation}
\rho\simeq \left(\frac{3^{n+1} h^{n+1}\tau}{A_{0}(t_{0}-t)^{n+1}}\right)^{\frac{1}{\alpha}}\phantom{spacespacesp}\text{for }\alpha>1\label{rhoin2}\,. 
\end{equation}

Eq.(\ref{rhoin}) corresponds to the cases when $\omega$ is a constant. For $n=0,1$, we find Eq.(\ref{zig}) and Eq.(\ref{zurp}). When $\alpha=1$ and $n>1$, the fluid energy density diverges faster than $H^{2}$ ($\sim (t_{0}-t)^{-2}$) and the EOM (\ref{EOM1})-(\ref{EOM2}) become inconsistent on the Big Rip. We can say that fluids with $\omega$ constant and bulk viscosity proportional to $H^{n}$, where $n>1$, avoid the Big Rip in $f(R,G)$-gravity. The same happens in the presence of viscous fluids with $\omega$ non constant and $n+1>2\alpha$, where $\alpha>1$, as in Eq.(\ref{rhoin2}).

For Type I singularities ($\beta>1$), an asymptotic, simple solution of Eq.(\ref{super}) is:
\begin{equation}
\rho\simeq \left(\frac{3^{n+1} h^{n+1}\tau}{A_{0}(t_{0}-t)^{(n+1)\beta}}\right)^{\frac{1}{\alpha}}\phantom{space}\text{for }\alpha\geqslant 1\label{rhoin3}\,.
\end{equation}

The cases $\alpha=1$ and $n=0,1$ correspond to Eq.(\ref{zagzag}) and Eq.(\ref{uno}). The fluid avoid the Type I singularities if $2\alpha<n+1$ when $\alpha\geqslant 1$, so that its energy density diverges faster than $H^{2}$ in Eq.(\ref{EOM1}). It means that, if the viscosity behaves as a power function of $H$ larger than one, the fluid is able to avoid the Big Rip and Type I singularities in $f(R,G)$-gravity. 

Note that the viscosity is introduced in the EOM by the fluid pressure of Eq.(\ref{eq.state}). On the Big Rip and Type I singularities, the curvature $R$ behaves as $H^{2}$ in Eq.(\ref{R}). Motivated by fact that the correction term $R^{\gamma}$ with $\gamma>2$ cures Big Rip and Type I singularities in $f(R,G)$ gravity\cite{Odintsov, miodue}, we may directly conclude that fluid viscosity proportional to $H^{\gamma-1}$ shows the same effect, like we have just seen.
    
For Type III singularities ($0<\beta<1$), an asymptotic solution of Eq.(\ref{super}) is: 
\begin{equation}
\rho\simeq \frac{3^{n+2} h^{n+2}\tau}{(2\beta+n\beta-1)(t_{0}-t)^{2\beta+n\beta-1}}\phantom{space}\text{for }1/2< \alpha\leqslant 1\label{rhoin4}\,.
\end{equation}

The cases $\alpha=1$ and $n=0,1$ correspond to Eq.(\ref{otto}) and Eq.(\ref{due}). The fluid energy density diverges faster than $H^{2}$ when $n>1/\beta$. In principle, if a $f(R,G)$-theory shows the Type III singularity for a certain value of $\beta$, the presence of a fluid with viscosity proportional to $H^{n}$, where $n>1/\beta$ (and, as a consequence, $n>1$), can make inconsistent the EOM and avoid this kind of singularity. Otherwise, it could appear a new Type III singularity realized by fluid for $H=h/(t_{0}-t)^{1/n}$, so that $\rho\sim H^{2}$, solving in some cases Eq.(\ref{EOM1}).

For Type II and IV singularities ($\beta<0$), if $H_{0}\neq 0$, an asymptotic solution of Eq.(\ref{super}) is given by: 
\begin{equation}
\rho\simeq \frac{3^{n+2}H_{0}^{n+1}h\tau}{(\beta-1)(t_{0}-t)^{\beta-1}}+\left(\frac{3^{n+1}H_{0}^{n+1}\tau}{A_{0}}\right)^{\alpha}\phantom{space}\text{for }\alpha\geqslant 1\label{rhoin5}\,.
\end{equation}

The cases $\alpha=1$ and $n=0,1$ correspond to Eq. (\ref{zag}) and Eq.(\ref{tre}). In general, this kind of fluid influences the feature of Type II and IV singularities in $f(R,G)$-gravity, but not necessarly avoid they.\\

If the viscosity is equal to zero ($\tau=0$), Eq.(\ref{omega}) and Eq.(\ref{super}) yield:
\begin{equation}
\rho=\left[(\alpha-1)\left(3A_{0}\ln \frac{a(t)}{a_{0}}\right)\right]^{\frac{1}{1-\alpha}}\,,
\end{equation}
where $a(t)$ is, as usual, the scale factor, $a_{0}$ a positive parameter and $\alpha\neq 1$ (non perfect fluids). We may take $\alpha>1$ and $A_{0}$ positive, and consider the parameter $a_{0}$ very small with respect to $a(t)$, so that, in general, $\rho$ is positive. 
 
In addition, we set
\begin{equation}
(3A_{0}(\alpha-1))^{\frac{1}{1-\alpha}}=\frac{H_{0}^{2}}{8 \pi G_{N}}\,,\label{Lambda}
\end{equation}
where $H_{0}$ is a positive parameter. As a consequence, we have:
\begin{equation}
\rho=\frac{H_{0}^{2}}{8 \pi G_{N}}\left[\ln \frac{a(t)}{a_{0}}\right]^{\frac{1}{1-\alpha}}\,.\label{Lorenzo}
\end{equation}

In Einstein's gravity ($f(R,G)=0$), the first equation of motion (\ref{EOM1}) reads:
\begin{equation}
a(t)=a_{0}\text{Exp}\Bigl\{ 6^{\frac{2-2\alpha}{2\alpha-1}}\left[\pm \dfrac{(2\alpha-1)(\sqrt{3}H_{0} t)}{\alpha-1}\right]^{2(\alpha-1)/(2\alpha-1)}\Bigr\}\label{zirp}\,.
\end{equation}

Note that for large values of $\alpha$, the fluid energy density tends to $H_{0}^{2}/8\pi G_{N}$, and Eq.(\ref{omega}), by using Eq.(\ref{Lambda}), leads to $\omega\simeq -1$, and $a(t)\simeq a_{0}e^{H_{0}t/3}$ ((anti)de-Sitter universe).
%
%

By matching Eq.(\ref{b}) and Eq.(\ref{zirp}), we can see that the fluid produces the Type I singularity for $\beta=1/(2\alpha-1)$, where $1/2<\alpha<1$, and in general does not avoid Big Rip and Type I singularities in $f(R,G)$-gravity.

Since on Type III singularity the fluid energy density tends to a constant  in Eq.(\ref{Lorenzo}), it does not avoid this singularity in $f(R,G)$-gravity, when the effective energy density diverges.
 
At last, on Types II and IV singularities, it is easy to see that, by substituting Eq.(\ref{c}) into the scale factor $a(t)$, the asymptotic time-dependent part of Eq.(\ref{Lorenzo}) behaves as $\sim(t_{0}-t)$ and it is larger than the time-dependent part of $H^{2}$ ($\sim(t_{0}-t)^{-\beta}$), when $\beta<-1$, in Eq.(\ref{EOM1}), so that, in general, especially Type IV singularities become difficoult to realize in $f(R,G)$-gravity in the presence of this kind of fluids.\\

A special case of non-viscous fluid with $\omega$ non-constant is the Chaplygin gas\cite{Ciappinski}, whose equation of state is:
\begin{equation}
p=-\frac{A_{0}}{\rho}\,,
\end{equation}
where $A_{0}$ is a positive constant. Eq.(\ref{conservationlaw}) leads:
\begin{equation}
\rho=\sqrt{A_{0}+\frac{1}{a(t)^{6}}}\label{Todt}\,.
\end{equation}

Since $a(t)$ does not tend to zero for any type of singularities, the energy density always tends to a constant in the limit $t\rightarrow t_{0}$, and the Chaplygin gas does not influence the asymptotic behaviour of $f(R,G)$-models in wich appear Big Rip or Type I and III singularities. In the case of $f(R,G)$-models that realize Type II and IV singularities, the energy density is not negligible and could influence the behaviur of the EOM (\ref{EOM1})-(\ref{EOM2}), avoiding Type IV singularities.\\
\\
Example:
\begin{itemize}

\item In the model $f(R)=\alpha R^{1/2}$, with $\alpha$ positive constant, the Type III singularity for $\beta=1/3$ could appear  \cite{miodue}. A fluid with $\omega$ constant and viscosity proportional to $H^{n}$, where $n>3$, avoid this kind of singularity. It is interesting to see that in this case, since if $\beta=1/n$, $\rho_{G}$ diverges faster than $H^{2}$, the fluid does not produce a new Type III singularity, due to the contribute of modified gravity. Moreover, the model is free of any other type of singularity.  
 
\end{itemize}

\setcounter{equation}{0}

\section{Quantum effects} 

\paragraph*{} Consider next the quantum contribution to the conformal anomaly. The complete fluid energy density is:
\begin{equation}
\rho_{tot}=\rho+\rho_{A}\,,
\end{equation}
where $\rho_{A}$ is given by quantum effects. Taking the trace of the conformal anomaly energy-momentum tensor,
\begin{equation}
T_{A}=-\rho_{A}+3p_{A}\,,
\end{equation}
plus observing the energy conservation law,
\begin{equation}
\dot{\rho_{A}}+3H(\rho_{A}+p_{A})=0\,,
\end{equation}
we find that:
\begin{equation}
p_{A}=-\rho_{A}-\frac{\dot{\rho_{A}}}{3H}\,.
\end{equation}

Thus we obtain for the conformal anomaly energy density\cite{conformalanomaly}:
\begin{eqnarray}
& & \phantom{spacespace}\rho_{A}=-\frac{1}{a^{4}}\int dt a^{4}H T_{a}=\nonumber\\ \nonumber \\
& & -\frac{1}{a^{4}}\int dt a^{4}H\Bigl\{-12b\dot{H}^{2}+24b'(-\dot{H}^{2}+H^{2}\dot{H}+H^{4})-\nonumber\\ \nonumber \\ 
& & \phantom{spacespace}(4b+6b'')(\dddot{H}+7H\ddot{H}+4\dot{H}^{2}+12H^{2}\dot{H})\Bigr\}\,.\label{equazione}
\end{eqnarray}
Here, $b$, $b'$ and $b''$ are constants, occuring in the expression for the conformal trace anomaly:
\begin{equation}
T_{A}=b(F+\frac{2}{3}\Box R)+b'G+b''R\,.
\end{equation}
Here, $F$ is the squared Weyl Tensor and $G$ the Gauss-Bonnet invariant. Explicitly, if there are $N$ scalars, $N_{1/2}$ spinors, $N_{2}$ gravitons and $N_{HD}$ higher derivative conformal scalars, one has for $b$ and $b'$ the following expressions:
\begin{equation}
b=\frac{N+6N_{1/2}+12N_{1}+611N_{2}-8N_{HD}}{120(4\pi)^{2}}\,,
\end{equation}
\begin{equation}
b'=\frac{N+11N_{1/2}+62N_{1}+1411N_{2}-28N_{HD}}{360(4\pi)^{2}}\,,
\end{equation}
whereas $b''$ is an arbitrary constant whose value depends on the regularization.

The quantum corrected FRW Eq.(\ref{EOM1}) is:
\begin{equation}
\rho_{G}+\rho+\rho_{A}=\frac{3}{8\pi G_{N}}H^{2}\label{EOM1corrected}\,.
\end{equation}

Quantum effects become relevant for large values of curvature $R$ and when the effective energy density of the universe is not too much large. In particular, this is the case of Type II singularities. 
\\

Let us consider a simple example of realistic $f(R)$-model. The Hu-Sawicki Model\cite{HuSaw} reproduces the current acceleration of universe:
\begin{equation}
f(R)=-\frac{m^{2}c_{1}(R/m^{2})^{n}}{c_{2}(R/m^{2})^{n}+1}=-\frac{m^{2}c_{1}}{c_{2}}+\frac{m^{2}c_{1}/c_{2}}{c_{2}(R/m^{2})^{n}+1}\,,
\end{equation}
where $m^{2}$ is a mass scale, $c_{1}$ and $c_{2}$ are positive parameters and $n$ a natural number. 

This is a ``one-step model'' (for reviews, see \cite{Starobinski, mio}) and it is characterized by the existence of one transition scalar curvature, so that, for $R=0$ the modified gravity vanishes and we find the Minkowski solution of Special Relativity, and, for large value of $R$, the model exibits a constant asymptotic behaviour, namely $f(R>>m^{2})\simeq -c_{1}m^{2}/c_{2}$. The model has been very carefully constructed and $c_{1}m^{2}/c_{2}\simeq2\Lambda$, where $\Lambda$ is the Cosmological Constant.

The Hu-Sawiki Model could become singular. In particular, it shows a Type II singularity when $H$ behaves as:    
\begin{equation}
H=\frac{h}{(t_{0}-t)^{\beta}}+H_{0}\,,\phantom{sp}-1<\beta<0\,.
\end{equation}

By solving the asymptotic limit of equations of motion (\ref{EOM1bis})-(\ref{EOM2bis}) in the absence of fluids, when $\beta=-n/n+2$, we find:
\begin{equation}
h= \left(\frac{6n^{2}(n+1)}{(n+2)^{2}}\left (\frac{2+n}{-6n}\right)^{n+2}\left(\frac{c_{1}}{c_{2}^{2}}\phantom{s}m^{2(n+1)}\right)\right)^{n+2}\,,
\end{equation}
\begin{equation}
H_{0}=\sqrt{\frac{c_{1}m^{2}}{6c_{2}}}\label{accazero}\,.
\end{equation}
Here, $h$ is positive if $n$ is an even number. In this case the model could show the Type II singularity. Note that $H_{0}=H_{dS}$, where $H_{dS}$ is the constant Hubble parameter in the de Sitter universe, because the contribute to the constant derives from the term in $f(R)$ into Eq.(\ref{rhoG}), and $f(R)$ tends to the Cosmological Constant for large positive or negative values of $R$. Not necessarly a one step model tends to the Cosmological Constant for negative values of $R$ (see, for instance, the exponential models in Ref.\cite{mio}).

In order to avoid the singualrity, we can introduce, for example, a quintessence fluid in the cosmological scenario. On the Type II singularity, the quantum effects of fluid become relevant and Eq.(\ref{equazione}) behaves as:
\begin{equation}
\rho_{A}\simeq \frac{\alpha}{(t_{0}-t)^{\beta+2}}\,.
\end{equation}
Here, $\alpha$ is a number. This term is very large with respect to the density of quintessence fluid, and diverges in Eq.(\ref{EOM1corrected}), so that the model is protected by singularities.

Note that, in the de Sitter universe, quantum effects are negligible and the de Sitter solution is asymptotically stable, due to the fact that the quintessence fluid disappears in expanding universe.
 
\section{Conclusion}

\paragraph*{} In the present paper, we have investigated the
finite-time future singularities in $f(G,R)$-gravity in the presence of viscous fluids. 
We have considered different fluids and the possibility to avoid the singularities.

In principle, perfect fluids may add singularities in $f(R,G)$-models or make some singularities stronger, but, due to introdution of bulk viscosity, all the four types of singularity could be removed from theory. Note that viscous fluid does not mean the modification of gravity. This is just some effective fluid added to gravity under consideration and it suggest a possible scenario to protect the theory from singularities. 

It is interesting to compare these results with Dark Matter effects to future 
singularities as described in Ref.\cite{N-O}, where 
it was found that coupling of singular forms of Dark Energy with Dark Matter may prevent only from 
Type II and Type IV future singularities.
The viscosity leads to high curvature correction terms as a function
of Hubble parameter and the effect of viscous fluid maybe more similar to $R^2$ term which may cure 
all types of singularities as it is first shown in Ref.\cite{MCB}.
In this respect, it would be of interest to study the combined effect of 
Dark Matter and viscous fluid to future singularities in modified gravity. 

In addition, we have discussed a possible way to resolve
the singularity problem, by taking into account quantum gravity effects, that become important in the case of Type II singularities.

\section*{Acknowledgments}
\paragraph*{}We thank Professor Sergei Odintsov and Professor Sergio Zerbini for comments and valuable suggestions on this work. L.S. thanks Dr. Enrico Fiorentini for suggestions. O.G. was supported by ESF programme New Trends and Applications of The 
Casimir Effect. The work is supported in part by INFN (Trento)-CSIC (Barcelona) exchange grant. 

\setcounter{equation}{0}
\section*{Appendix}

\appendix

\section{Energy conditions near the singularities}

\paragraph*{} We briefly discuss the energy conditions maybe related with occurrence of singularities. We have four types of energy conditions:
\begin{itemize}
\item Weak energy condition (WEC): $\rho\geqslant 0$ and $\rho+p\geqslant 0$; 
\item Strong energy condition (SEC): $\rho+p\geqslant 0$ and $\rho+3p\geqslant 0$;
\item Null energy condition (NEC): $\rho+p\geqslant 0$;
\item Dominant energy condition (DEC): $\rho\geqslant |p|$.
\end{itemize}
On the singular solution $H=h/(t_0-t)^{\beta}+H_{0}$, we have
\begin{equation}
\rho+p=-\frac{1}{4\pi G_{N}}\frac{h\beta}{(t_{0}-t)^{\beta+1}}\,,
\end{equation}
where $\rho$ and $p$ are the effective energy density and pressure of the universe (deriving from modified gravity and fluid).

The effective DE related with Type I and III singularities violate the NEC, whereas DE related with Types II and III satisfy the NEC.

Note that
\begin{equation}
 \rho+3p=-\frac{3}{4\pi G_{N}}\left(H_{0}^{2}+2\frac{h H_{0}}{(t_{0}-t)^{\beta}}+\frac{h^{2}}{(t_{0}-t)^{2\beta}}+\frac{h\beta}{(t_{0}-t)^{\beta+1}}\right)\,.\label{tatata}
\end{equation}

The effective DE related with Type II singularities ($-1<\beta<0$) violate the SEC for small value of $t$, but, when $t$ is close to $t_{0}$, the last term of Eq.(\ref{tatata}) is dominant and the SEC is satisfied on the singular solution.

Finally, in the case of Type IV singularities, when $t$ is really close to $t_{0}$, the term $H_{0}^2$ could be dominant and the SEC is violated, expecially if $|\beta|<<1$. 

At last, it is easy to see that, on the singular solutions, when $t$ is near to $t_{0}$, the DEC is always violated except for large value of $H_{0}$ in the case of Type IV singularities, but also in this case the behaviour of universe approaching the singular solution violate the DEC.

We note that in general viscous fluid curing the singularities not necessarly conflict with these energy conditions on singular solutions.


\end{document}